\documentclass[onecolumn]{IEEEtran}

\usepackage{psfrag,epsfig,graphics}
\usepackage{amsmath,amssymb,multirow}
\usepackage{subfigure}
\usepackage{mathtools}
\usepackage[below]{placeins}
\usepackage{cite}
\usepackage{algorithm}
\usepackage{algorithmic}

\newcommand{\cp}[1]{\ifmmode {\mathcal{#1}}\else ${\mathcal{#1}}$\fi}
\newcommand{\bzero}{\boldsymbol{0}}
\newcommand{\bone}{\boldsymbol{1}}

\newcommand{\bLambda}{\boldsymbol{\Lambda}}
\newcommand{\blambda}{\boldsymbol{\lambda}}
\newcommand{\ba}{\boldsymbol{a}}
\newcommand{\bA}{\boldsymbol{A}}
\newcommand{\bB}{\boldsymbol{B}}

\newcommand{\bC}{\boldsymbol{C}}

\newcommand{\bE}{\boldsymbol{E}}
\newcommand{\bEw}{\boldsymbol{E}_{\omega}}

\newcommand{\bI}{\boldsymbol{I}}
\newcommand{\bIw}{\boldsymbol{I}_{\omega}}

\newcommand{\bP}{\boldsymbol{P}}

\newcommand{\bQ}{\boldsymbol{Q}}

\newcommand{\br}{\boldsymbol{r}}
\newcommand{\bR}{\boldsymbol{R}}
\newcommand{\bs}{\boldsymbol{s}}
\newcommand{\bS}{\boldsymbol{S}}
\newcommand{\bSw}{\boldsymbol{S}_{\omega}}
\newcommand{\btS}{\boldsymbol{\widetilde{S}}}
\newcommand{\btSw}{\boldsymbol{\widetilde{S}}_\omega}
\newcommand{\bts}{\boldsymbol{\tilde{s}}}

\newcommand{\bv}{\boldsymbol{v}}

\newcommand{\bW}{\boldsymbol{W}}
\newcommand{\bx}{\boldsymbol{x}}

\newcommand{\bX}{\boldsymbol{X}}

\newcommand{\bz}{\boldsymbol{z}}
\newcommand{\bZ}{\boldsymbol{Z}}
 
\newcommand{\tr}{\text{trace}}

\title{\huge Blind and fully constrained unmixing \\ of hyperspectral images}

\author{\IEEEauthorblockN{Rita Ammanouil, Andr\'e Ferrari, C\'edric Richard, David Mary} \\
\IEEEauthorblockA{
Laboratoire Lagrange, Universit\'e de Nice Sophia-Antipolis, France \\
\{rita.ammanouil, andre.ferrari, cedric.richard, david.mary\}@unice.fr
}}

\begin{document}

\maketitle

\begin{abstract}
This paper addresses the problem of blind and fully constrained unmixing of hyperspectral images. Unmixing is performed without the use of any dictionary, and assumes that the number of constituent materials in the scene and their spectral signatures are unknown. The estimated abundances satisfy the desired sum-to-one and nonnegativity constraints. Two models with increasing complexity are developed to achieve this challenging task, depending on how noise interacts with hyperspectral data. The first one leads to a convex optimization problem, and is solved with the Alternating Direction Method of Multipliers. The second one accounts for signal-dependent noise, and is addressed with a Reweighted Least Squares algorithm. Experiments on synthetic and real data demonstrate the effectiveness of our approach. 
\end{abstract}

\newpage

\section{Introduction} 
\label{sec:intro}

Hyperspectral imaging is a continuously growing area of remote sensing, which has received considerable attention in the last decade. Hyperspectral data provide spectral images over hundreds of narrow and adjacent bands, coupled with a high spectral resolution. These characteristics are suitable for detection and classification of surfaces and chemical elements in the observed images. Applications include land use analysis, pollution monitoring, wide-area reconnaissance, and field surveillance, to cite a few. When unmixing hyperspectral images~\cite{Keshava02}, two types of pixels can be distinguished: the pure pixels and the mixed ones. Each pure pixel, also called endmember, contains the spectral signature of a constituent material in the scene, whereas a mixed pixel consists of a mixture of the endmembers. The fraction of each endmember in a mixed pixel is called abundance. Three consecutive tasks are usually required for unmixing: determining the number of endmembers, extracting the spectral signature of the endmembers, and estimating their abundances for every pixel in the scene. Several algorithms have been proposed to perform each stage separately. Virtual Dimensionality (VD) \cite{Chang04}, followed by N-FINDR \cite{Winter99} and FCLS \cite{Heinz01} is among the most widely used processing pipeline. Alternative methods jointly performs (part of) these tasks in order to solve the blind source separation problem \cite{Eches10,Honeine12,Miao07,Yang11}.

In order to introduce our approach, we shall now describe the noise-free case first. Consider the linear mixing model where a mixed pixel is expressed as a linear combination of the endmembers weighted by their fractional abundances. In matrix form, we simply have:
\begin{equation} 
	\label{LinearModel}
	\btS = \bR{\bA}
\end{equation}
where $\btS = (\bts_{1},\ldots, \bts_{N}) $, $\bR = (\br_{1},\ldots, \br_{M})$,  ${\bA} = ({\ba}_{1},\ldots, {\ba}_{M})^{\top}$, $\bts_{j}$ is the  $L$-dimensional  spectrum of the $j$-th pixel, $L$ is the number of frequency bands, $\br_{i}$ is ${L}$-dimensional  spectrum of the $i$-th endmember, $M$ is the number of endmembers, ${\ba}_i$ is the ${N}$-dimensional abundance map of the $i$-th endmember, and $N$ is the number of pixels in the image. Model \eqref{LinearModel} means that the $(i,j)$-th entry ${\bA}_{ij}$ of matrix $\bA$ represents the abundance of the endmember $\br_{i}$ in pixel $\bts_{j}$. The abundances obey the nonnegativity and sum-to-one constraints: $\bA_{ij} \geq 0 $ for all $i$ and $j$, and $ \sum_{i=1}^M \bA_{ij} = 1$ for all $j$. Note that the  tilde placed over symbols refers to noise-free data and all vectors are column vectors. 

In this study, we shall assume that the endmembers are unknown but present in the scene. Let $\omega$ be a subset of $N'$ indexes in $\{1, \ldots,N\}$ that contains at least the column index of each endmember. Under these assumptions, and without loss of generality, we observe that the mixing model \eqref{LinearModel} can be reformulated as follows
\begin{equation} 
	\label{LinearModel1}
	\btS = \btS_{\omega}\bX
\end{equation}
where $\btSw = (\bts_{{\omega}_1},\ldots, \bts_{{\omega}_{N'}}) $ denotes the restriction of $\btS$ to its columns indexed by  $\omega $, and ${\bX} = ({\bx}_{1},\ldots, {\bx}_{N'})^{\top}$ is the abundance matrix. Similarly as above, $\bX_{ij}$ is the abundance of $\bts_{{\omega}_i}$ in $\bts_j$. On the one hand, if $\bts_{\omega{_i}}$ is an endmember, ${\bx}_i$ has non-zero entries and represents the corresponding abundance map. On the other hand, if $\bts_{\omega{_i}}$ is a mixed pixel, ${\bx_i}$ has all its elements equal to zero. As a consequence, $\bX$ admits $N'-M$ rows of zeros, the other rows being equal to rows of $\bA$. This means that $\bX$ allows to identify the endmembers in $\btS$ through its non-zero rows, which is an interesting property to be exploited in the case where the endmembers are unknown. 
Let us now turn to the more realistic situation where some noise corrupts the observations. In this case, model \eqref{LinearModel1} becomes
\begin{equation} \label{LinearModel2}
\bS = \btS + \bE = \btS_{\omega}\bX + \bE
\end{equation}
where $\bS$ denotes the available data and $\bE$ the noise.

The aim of this paper is to derive two unmixing approaches with increasing complexity, depending on how noise is to be handled. These methods are blind in the sense that the endmembers and their cardinality are unknown. The first one considers the approximate model
\begin{equation}
	\label{LinearModel3}
	\bS \approx \bS_{\omega}\bX + \bE
\end{equation}
Compared to \eqref{LinearModel2}, we thus assume that noise does not dramatically affect factorization of the mixing process, which is valid for very high signal-to-noise ratio (SNR).
With this approach, we shall look for a few columns of $\bS_{\omega}$ that can effectively represent the whole scene. This strategy subserves a blind and self-dependent framework. It departs from methods based on a preselected dictionary of endmembers estimated from other experimental conditions, and thus do not accurately represent the endmembers in $\bR$. In order to estimate the abundance matrix $\bX$, we use prior information. First, we impose that the estimated abundances obey the non-negativity and sum-to-one constraints, namely, $\bX_{ij} \geq 0 $ for all $(i,j)$, and $ \sum_{i=1}^N \bX_{ij} = 1$ for all $j$.  In addition, as discussed above, the algorithm has to force rows of $\bX$ to be zero vectors in order to identify the endmembers. Because the locations and the cardinality of the endmembers are unknown, the set of candidates has to be sufficiently large, that is, $N' \gg M$. We thus expect many rows in $\bX$ to be equal to zero. To promote this effect, the so-called Group Lasso $\ell_{2,1}$-norm regularization can be employed \cite{Yuan06}. Because model~\eqref{LinearModel3}  becomes a poor approximation of model~\eqref{LinearModel2} as the noise power increases, we shall also propose an alternative strategy to solve the unmixing problem based on the exact model \eqref{LinearModel2}. The first approach leads to a convex optimization problem that can be solved with the Alternating Direction Method of Multipliers (ADMM)~\cite{Boyd10}. The second one takes the noise in $\bS_{\omega}$ into account, which results in a non-convex and heteroscedastic optimization problem. The latter will be solved with an Iterative Reweighted Least Squares (IRLS)  algorithm. 

Few models of the form~\eqref{LinearModel2} have been studied in the literature \cite{Esser12,Fu13,Iordache13}. These last three works assume that $\bS_{\omega}$ is noise-free. Moreover, in \cite{Esser12}, the authors use an $\ell_\infty$-norm rather than the $\ell_{2,1}$-norm regularization considered here. In \cite{Fu13}, the authors derive a Matching Pursuit approach \cite{Mallat93} in order to estimate the endmembers. A similar technique is considered in \cite{Iordache13}, but the authors do not assume that the endmembers are present in the scene and use a predefined dictionary. Finally, note that the non-negativity and sum-to-one constraints are not considered in \cite{Esser12,Fu13}. 

The rest of this paper is organized as follows. Sections \ref{sec:algGLUP} and \ref{sec:NGLUP} respectively describe the unmixing models \eqref{LinearModel3} and \eqref{LinearModel2}, and the corresponding estimation methods. Section \ref{sec:exp} provides experimental results on synthetic and real data. Finally, Section \ref{sec:conclusion} concludes this paper.

\section{Group Lasso with Unit sum and Positivity constraints (GLUP)} 
\label{sec:algGLUP}
\subsection{Model description}
The aim of this section is to derive the estimation method for model~\eqref{LinearModel3}, and finally define each step of the ADMM run to get the solution. In this approximate model, we assume that the noise $\bE$ is Gaussian independent and identically distributed, with zero mean and possibly unknown variance $\sigma^2$, that is, $\bE_{k,i}\sim\mathcal{N}(0,\sigma^2)$. The negative log-likelihood for model \eqref{LinearModel3} is given by
\begin{equation}
	\mathcal{L}(\bX) =\frac{NL}{2}\log (2\pi) +\frac{NL}{2}\log (\sigma^2) + \frac{1}{2\sigma^2} \| \bS-\bS_{\omega}\bX \|^2_F  \label{NegLogLik1}
\end{equation}
The Maximum Likelihood (ML) estimate, namely, the minimizer of $\mathcal{L}(\bX)$, is the solution of the usual Least Squares (LS) approximation problem $\min_{\bX}\|\bS-\bS_{\omega}\bX \|^2_F$. Since model \eqref{LinearModel3} follows from an approximation of model \eqref{LinearModel2}, the relevance of this LS fidelity term is essentially to ensure  that $\bS_{\omega}\bX$ matches $\bS$. The unmixing problem under investigation, however, requires that $\bX$ only has a few rows different from zero, in addition to the non-negativity and sum-to-one constraints. This leads to following convex optimization problem
\begin{equation} \label{cvx1}
	\begin{array}{ll} 
	\min_{\bX}   	& \frac{1}{2}\|\bS-\bS_{\omega}\bX\|_\text{F}^2+ \mu \sum_{k=1}^N  \| \bx_{k}\|_2  \\
	\text{subject to} & \bX_{ij} \geq 0 \quad \forall\, i,j \\ 
			&  \sum_{i=1}^N \bX_{ij} = 1 \quad \forall\, j
	\end{array}
\end{equation} 
with $\mu \geq 0$ a regularization parameter and $\bx_k$ the $k$-th row of $\bX$. The Group Lasso regularization term induces sparsity in the estimated abundance matrix at the group level \cite{Yuan06}, by possibly driving all the entries in several rows $\bx_k$ of $\bX$ to zero.  It is worth noting that when $\mu =0$ and $\bS_{\omega}=\bS$, the solution of problem \eqref{cvx1} is the identity matrix $\bX=\bI$. It follows that the efficiency of the approach relies on the $\ell_{2,1}$-norm regularization function.


\subsection{ADMM algorithm}
The solution of problem \eqref{cvx1} can be obtained in a simple and flexible manner using the ADMM algorithm~\cite{Boyd10}. We consider the canonical form
\begin{equation} \label{cvx2}
	\begin{array}{ll} 
	\min_{\bX,\bZ}   	& \frac{1}{2}\|\bS-\bS_{\omega}\bX\|_\text{F}^2+ \mu \sum_{k=1}^N  \|\bz_{k}\|_2 +  \cp{I}(\bZ)  \\
	\text{subject to} 		&  \bA\bX + \bB\bZ = \bC 
	\end{array}
\end{equation} 
with
\begin{equation}
	\bA = \left(\begin{array}{c} \bI\, \\ \bone^ \top \end{array}\right),\;
	\bB = \left(\begin{array}{c} -\bI \\ \,\bzero^\top\end{array}\right),\;
	\bC = \left(\begin{array}{c} \bzero\, \\  \bone^\top  \end{array}\right),  \nonumber
\end{equation}
where $\cp{I}$ is the indicator of the positive orthant guarantying the positivity constraint, that is, $\cp{I}(\bZ)= 0$ if $\bZ\succeq\bzero$ and $+\infty$ otherwise. The equality constraint imposes the consensus $\bX=\bZ$ and the sum-to-one constraint. In matrix form, the augmented Lagrangian for problem \eqref{cvx2} is given by \cite{Ekstein92} 
\begin{multline*}
	\mathcal{L}_\rho(\bX,\bZ,\bLambda) = \frac{1}{2}\|\bS-\bS_{\omega}\bX\|_F^2+ \mu \sum_{k=1}^N  \|\bz_{k}\|_2  +\cp{I}(\bZ)+ 
	\tr(\bLambda^\top (\bA\bX + \bB\bZ - \bC))+
	\frac{\rho}{2}\, \| \bA\bX + \bB\bZ - \bC\|_\text{F}^2
\end{multline*}
where $\bLambda$ is the matrix of Lagrange multipliers, $\mu$ and $\rho$ are positive regularization and penalty parameters, respectively. The flexibility of the ADMM lies in the fact that it splits the initial variable $\bX$ into two variables, $\bX$ and $\bZ$, and equivalently the initial problem into two subproblems. At iteration $k+1$, the ADMM algorithm is outlined by three sequential steps:

\subsubsection{Minimization of $\mathcal{L}_\rho(\bX,\bZ^k,\bLambda^k)$ with respect to $\bX$}
This step takes into account the previous estimates of $\bZ$ and $\bLambda$. The augmented Lagrangian is quadratic in terms of $\bX$. As a result, the solution has an analytical expression that is obtained by setting the gradient of $\mathcal{L}_\rho(\bX,\bZ^k,\bLambda^k)$ to zero:
\begin{equation}
	\bX^{k+1} = ({\bS_{\omega}}^\top\bS_{\omega} +\rho \bA^\top\bA)^{-1}(\bS_{\omega}^\top\bS - \bA^\top [\bLambda^k+\rho\,(\bB\bZ^k-\bC)]) 
\end{equation}

\subsubsection{Minimization of $\mathcal{L}_\rho(\bX^{k+1},\bZ,\bLambda^k)$ with respect to $\bZ$}
\label{subsec:zminstep}
After removing the terms that are independent of $\bZ$, the minimization of $L_\rho(\bX^{k+1},\bZ,\bLambda^k)$ with respect to $\bZ$ reduces to solving the following problem:
\begin{equation}
	\label{ADMMz}
	\begin{array}{ll} 
		\min_{\bZ}   & \mu \sum_{k=1}^N  \|\bz_{k}\|_2 +\tr(\bLambda^\top \bB\bZ)+\frac{\rho}{2} \| \bA\bX + \bB\bZ - \bC\|_\text{F}^2\\
		\text{subject to} & \bZ\succeq\bzero
	\end{array}
\end{equation}
This minimization step can be split into $N$ problems given the structure of matrices $\bA$ and $\bB$, one for each row of $\bZ$, that is,
\begin{equation}
	\label{zmin}
	\begin{array}{ll} 
		\min_{\bz}   & \frac{1}{2} \|\bz-\bv\|_2^2 + \alpha  \|\bz\|_2 + \cp{I}(\bz)
	\end{array}
\end{equation} 
where $\bv = \bx +\rho ^{-1} \blambda$, $\alpha = \rho ^{-1}\mu$, $\blambda$, $\bx$ and $\bz$ correspond to a row in $\bLambda$, $\bX$ and $\bZ$ respectively. The minimization problem~\eqref{zmin} admits a unique solution, given by the proximity operator \cite{Combettes09} of $f(\bz)=\alpha  \|\bz\|_2  + \cp{I}(\bz)$
\begin{equation}
	\label{Pmisto}
	\left\{
	\begin{array}{ll}
		\bz^\ast = \bzero  & \text{if } \|(\bv)_+\|_2 < \alpha  \\
		\bz^\ast  = \left(1 - \frac{\alpha}{\|(\bv)_+\|_2}\right) (\bv)_+  & \text{otherwise} 
	\end{array}\right.
\end{equation}
where $(\cdot)_+ = \max(\bzero,\cdot)$. On the one hand, the proximity operator of $f_1(\bz)=\alpha  \|\bz\|_2$  is the \emph{Multidimensional Shrinkage Thresholding Operator} (MiSTO) \cite{Puig11}. On the other hand, the proximity operator of the indicator function $f_2(\bz)=\cp{I}(\bz)$ is the projection onto the positive orthant. The proximity operator of $f(\bz)$ in \eqref{Pmisto}, that we refer  to as Positively constrained MiSTO, is an extension of both previous operators. The solution is of the form $\text{prox}_{f}=\text{prox}_{f_1}\circ \text{prox}_{f_2}$, that is, the thresholding of the projection. Operator \eqref{Pmisto} was recently used in \cite{Thiebaut13}. A proof for this operator can be found in the Appendix. 

\subsubsection{Update of the Lagrange multipliers $\bLambda$}
\label{subsec:updatestep}
Update of the Lagrange multipliers is carried out at the end of each iteration. $\bLambda^{k+1}$ represents the running sum of residuals. It gives an insight on the convergence of the algorithm. As $k$ tends to infinity, the primal residual tends to zero and $\bLambda^{k+1}$ converges to the dual optimal point. 
\begin{equation}
	\label{ADMML}
	\bLambda^{k+1} = \bLambda^{k} +\rho(\bA\bX^{k+1}+\bB\bZ^{k+1}-\bC).
\end{equation}
As suggested in \cite{Boyd10}, a reasonable stopping criteria is that the primal and dual residuals must be smaller than some tolerance thresholds, namely,
\begin{equation}
\|\bA\bX^{k+1}+\bB\bZ^{k+1}-\bC\|_2 \leq \epsilon_{\text{pri}} \quad \text{and} \quad \|\rho\bA^\top\bB(\bZ^{k+1}-\bZ^k)\|_2 \leq \epsilon_{\text{dual}}
\end{equation}

The pseudocode for the so-called GLUP method is provided by Algorithm \ref{pseudoGLUP}. It is worth emphasizing that the main difference between the ADMM steps developed in GLUP and those in \cite{Iordache13} arises in the ADMM variable splitting. The global problem in \cite{Iordache13} is decomposed into three subproblems: the least squares minimization, the Group Lasso regularization, and projection on the positive orthant. A consequence is that three ADMM variables are used instead of two, which leads to additional steps. In addition, the sum-to-one constraint is not considered in \cite{Iordache13}.

\medskip

\begin{algorithm*}
\caption{ : $\bX = \text{GLUP}(\bS,\bS_{\omega},\rho,\mu)$}
\label{pseudoGLUP}
\begin{algorithmic}[1]
\medskip\STATE Precompute $\bA$, $\bB$, and $\bC$
\STATE Initialize $\bZ = \bzero$ and $\bLambda= \bzero$
\STATE $\bQ = ({\bS_{\omega}}^\top\bS_{\omega} +\rho \bA^\top\bA)^{-1}$
\WHILE{  $\|\bR\|_2 \geq  \epsilon_{\text{pri}}$ or  $\|\bP\|_2 \geq  \epsilon_{\text{dual}}$ }
\STATE $\bX = \bQ(\bS_{\omega}^\top\bS - \bA^\top (\bLambda+\rho\,[\bB\bZ-\bC]))$
\STATE $\bZ^{\text{old}} = \bZ$
\FOR{$i=1\cdots N'$}
\STATE $\bv_i = ((\bx_i)^{\top} +\rho ^{-1} \blambda_i)_{+}$
\IF {$ \|\bv_i\|_2 < \rho ^{-1}\mu $}
\STATE $ \bz_i= \bzero $
\ELSE
\STATE $\bz_i = \left(1 - \frac{\mu}{\rho\|\bv_i\|_2}\right) \bv_i $
\ENDIF
\ENDFOR
\STATE $\bR = \bA\bX + \bB\bZ - \bC$
\STATE $\bP = \rho\bA\bB(\bZ-\bZ_{\text{old}})$
\STATE $ \bLambda = \bLambda +\rho(\bA\bX+\bB\bZ-\bC) $ 
\ENDWHILE
\end{algorithmic}
\end{algorithm*}

\section{Reduced Noise For Group Lasso with Unit Sum and Positivity constraints (NGLUP)}
\label{sec:NGLUP}
\subsection{Model description}
We now turn to the more realistic model \eqref{LinearModel2}. Let $\bEw$ and $\bIw$ be the $L$-by-$N'$ and $N$-by-$N'$ restrictions of $\bE$ and $\bI$ to the columns indexed by $\omega$, respectively. The noisy mixing model \eqref{LinearModel2} is given by
\begin{equation}
	\bS = (\bSw - \bEw)\bX + \bE = \bS_{\omega}\bX +\bE(\bI-\bIw\bX) 
	\label{model2}
\end{equation}
This model belongs to the family of heteroscedastic regression \cite{Hopper93}, where the variance of the additive noise depends on $\bX$. Let us define the matrix $\bC(\bX)$ as
\begin{equation}
	\label{C_exp}
	\bC(\bX)=(\bI - \bIw\bX)^{\top}(\bI - \bIw\bX)
\end{equation}
It follows that
\begin{equation}
\text{vec}(\bE(\bI-\bIw\bX))\sim \mathcal{N}(\bzero,\sigma^2\bC(\bX)\otimes \bI)
\end{equation}
where $\otimes$ represents the Kronecker product of matrices, and $\text{vec}(\cdot)$ is the operator that stacks the columns of a matrix on top of each other. The presence of $\bX$ in the expression of the noise variance has consequences on the negative log-likelihood of model $\eqref{model2}$, which no longer leads to the LS approximation problem
\begin{equation}
	\label{NegLogLik}
	\begin{split}
		\mathcal{L}(\bX,\sigma^2) &= \frac{1}{2}\log |\sigma^2\bC(\bX)\otimes \bI | 
			+\frac{1}{2} \text{vec}(\bS-\bS_{\omega}\bX)^{\top}(\sigma^2\bC(\bX)\otimes \bI)^{-1} \text{vec}(\bS-\bS_{\omega}\bX)  \\
		&=\frac{L}{2}\log |\sigma^2\bC(\bX)| + \frac{1}{2}  \tr ((\bS-\bS_{\omega}\bX) (\sigma^2\bC(\bX))^{-1} (\bS-\bS_{\omega}\bX)^{\top})  \\
		&=\frac{L}{2}\log |\sigma^2\bC(\bX)| + \frac{1}{2} \| \bS-\bS_{\omega}\bX \|^2_{(\sigma^2\bC(\bX))^{-1}}  
	\end{split}
\end{equation}

The ML estimate for problem \eqref{NegLogLik} with the Group Lasso regularization, nonnegativity and sum-to-one constraints yields the following constrained optimization problem:
\begin{equation}
	\label{noncvx1}
	\begin{array}{ll} 
	\min_{\bX, \sigma^2}   	& \frac{L}{2}\log |\sigma^2\bC(\bX)| + \frac{1}{2} \| \bS-\bS_{\omega}\bX \|^2_{(\sigma^2\bC(\bX))^{-1}}  + \mu \sum_{k=1}^N  \| \bx_{k}\|_2  \\
	\text{subject to} &  \bX_{ij} \geq 0 \quad \forall\, i,j\\ 
			& \sum_{i=1}^N \bX_{ij} = 1 \quad  \forall\, j
	\end{array}
\end{equation}

\subsection{Alternating ADMM algorithm}
Problem \eqref{noncvx1} is not convex and requires the estimation of $\sigma^2$. The second term in the objective function is closely related to Iteratively Reweighted Least Squares (IRLS) algorithms used as a solution in heteroscedastic models \cite{Daubechies10}. Note that, in IRLS algorithms, $(\bS-\bS_{\omega}\bX) $ in equation \eqref{NegLogLik} is usually substituted by $(\bS-\bS_{\omega}\bX) ^{\top}$. This has consequences on the $\bX$ minimization step. In IRLS, the estimation process is carried out in two steps. The first step consists of updating weights, which are usually set to be inversely proportional to variances. The second step is the calculation of the LS estimator using the updated weights. Many strategies can be used to estimate the variances for the weight matrix, see for example \cite{Wayne78, Caroll88 , Hopper93}. 

The resolution of problem \eqref{noncvx1} with respect to $\sigma^2$ for fixed $\bX$ gives
\begin{equation}
\label{sigma_exp}
		\sigma^2(\bX)= \frac{1}{NL}\tr ((\bS-\bS_{\omega}\bX)\,\bC(\bX)^{-1} (\bS-\bS_{\omega}\bX)^{\top})
\end{equation}
Let $\bW(\bX) = \sigma^2(\bX)\bC(\bX)$ denote the weight matrix of the least squares term in \eqref{noncvx1}. To solve problem \eqref{noncvx1} with respect to $\sigma^2$ and $\bX$, we propose to proceed iteratively. Let $\bX^{k}$ be the solution of the previous iteration. The first step consists of calculating $\bW(\bX^k)$ using equations \eqref{C_exp} and \eqref{sigma_exp}. In the second step, this updated weight matrix is used to estimate $\bX^{k+1}$ as follows
\begin{equation}
	\label{noncvx2}
	\begin{array}{ll} 
	\min_{\bX}   	& \frac{1}{2} \| \bS-\bS_{\omega}\bX \|^2_{(\bW^k)^{-1}} + \mu \sum_{k=1}^N  \| \bx_{k}\|_2  \\
	\text{subject to} & \bX_{ij} \geq 0 \quad \forall\,i,j \\ 
			         & \sum_{i=1}^N \bX_{ij} = 1 \quad \forall\, j
	\end{array}
\end{equation}
where $\bW^k = \bW(\bX^k)$. Given  $\bW^k$, problem \eqref{noncvx2} reduces to a weighted version of GLUP \eqref{cvx1} due to the weighted norm in the first term. The ADMM solution developed in section \ref{sec:algGLUP} can be adapted to solve the optimization problem~\eqref{noncvx2}. Minimizing the augmented Lagrangian with respect to $\bZ$, and updating the Lagrange multipliers, can by carried out exactly as in Section~\ref{sec:algGLUP}. For concision, only the $\bX$-minimization step is described hereafter. 

\subsubsection*{Minimization of $\mathcal{L}_\rho(\bX,\bZ^k,\bLambda^k)$ with respect to $\bX$}
Omitting the terms that do not depend on $\bX$, the minimization of the augmented Lagrangian $\mathcal{L}_\rho(\bX,\bZ^k,\bLambda^k)$  with respect to $\bX$ leads to
\begin{equation}
\label{xmin2}
	\begin{array}{ll} 
	\min_{\bX}   	& \frac{1}{2} \| \bS-\bS_{\omega}\bX\|^2_{(\bW^k)^{-1}} +\tr(\bLambda^\top (\bA\bX)) +\frac{\rho}{2} \| \bA \bX+\bB \bZ-\bC\|^2_{\text{F}} 
         \end{array}
\end{equation}
Problem \eqref{xmin2} is quadratic in $\bX$ and admits an analytical solution obtained by setting the gradient to zero. This amounts to solving the Sylvester equation, which has an analytic solution \cite{Bartels72}
\begin{equation}
\label{sylvester}
	\bS_{\omega}^{\top}\bS_{\omega}\bX(\bW^k)^{-1} +\rho \bA^{\top}\bA\bX = \bS_{\omega}^{\top}\bS(\bW^k)^{-1} - \rho \bA^{\top}\left(\bB\bZ^k-\bC+\frac{\bLambda^k}{\rho}\right) 
\end{equation}

Problem \eqref{noncvx1} is not convex. An alternating optimization algorithm is more likely to converge to local minima with worse accuracy than the convex version. For this reason, we suggest, as a warm start, to initialize NGLUP with GLUP estimate. Algorithm \ref{alg:pseudoNGLUP} provides the pseudocode for NGLUP. The algorithm contains two main loops. The inner loop aims at finding the solution of problem \eqref{noncvx2}, whereas the outer loop updates the least-square weight matrix. 

\begin{algorithm}
\caption{ : $\bX = \text{NGLUP}(\bS,\bS_{\omega},\rho^\circ,\mu^\circ,\rho,\mu)$}
\label{alg:pseudoNGLUP}
\begin{algorithmic}[1]
\STATE Precompute $\bA$, $\bB$, and $\bC$
\STATE Initialize $\bX=\text{GLUP}(\bS,\bS_{\omega},\rho^\circ,\mu^\circ)$, $\bZ=\bX$, $\bLambda=\bzero$
\WHILE{ $\| \bX-\bX_{\text{old}} \|_2 \geq  \epsilon_\text{tol}$ }
\STATE $\bC(\bX)=(\bI - \bIw\bX)^{\top}(\bI - \bIw\bX)$
\STATE $\sigma^2(\bX)= \frac{1}{NL}\tr ((\bS-\bS_{\omega}\bX)\bC(\bX)^{-1} (\bS-\bS_{\omega}\bX)^{\top})$
\STATE $\bW(\bX)=\sigma^2(\bX)\bC(\bX)$
\STATE $\bX^{\text{old}}=\bX$, $J = 1$
\WHILE{ ($\|\bR\|_2 \geq  \epsilon_{\text{pri}}$ or  $\|\bP\|_2 \geq  \epsilon_{\text{dual}}$) and ($J \leq J_{\text{max}}$)}
\STATE $\bX = $ solution of Sylvester equation \eqref{sylvester}
\STATE $\bZ_{\text{old}}=\bZ$
\FOR{$i=1\cdots N'$}
\STATE $\bv_i = ((\bx_i)^{\top} +\rho ^{-1} \blambda_i)_{+}$
\IF {$ \|\bv_i\|_2 < \rho ^{-1}\mu $}
\STATE $ \bz_i= \bzero $
\ELSE
\STATE $\bz_i = \left(1 - \frac{\mu}{\rho\|\bv_i\|_2}\right) \bv_i $
\ENDIF
\ENDFOR
\STATE $\bR = \bA\bX + \bB\bZ - \bC$
\STATE $\bP = \rho\bA\bB(\bZ-\bZ_{\text{old}})$
\STATE $ \bLambda =  \bLambda +\rho(\bA\bX+\bB\bZ-\bC) $ 
\STATE $J = J + 1$
\ENDWHILE
\ENDWHILE
\end{algorithmic}
\end{algorithm}

\section{experimental results}
\label{sec:exp}
\subsection{Synthetic Data}
\begin{figure}[h]
    \centering
    \includegraphics[scale=0.4]{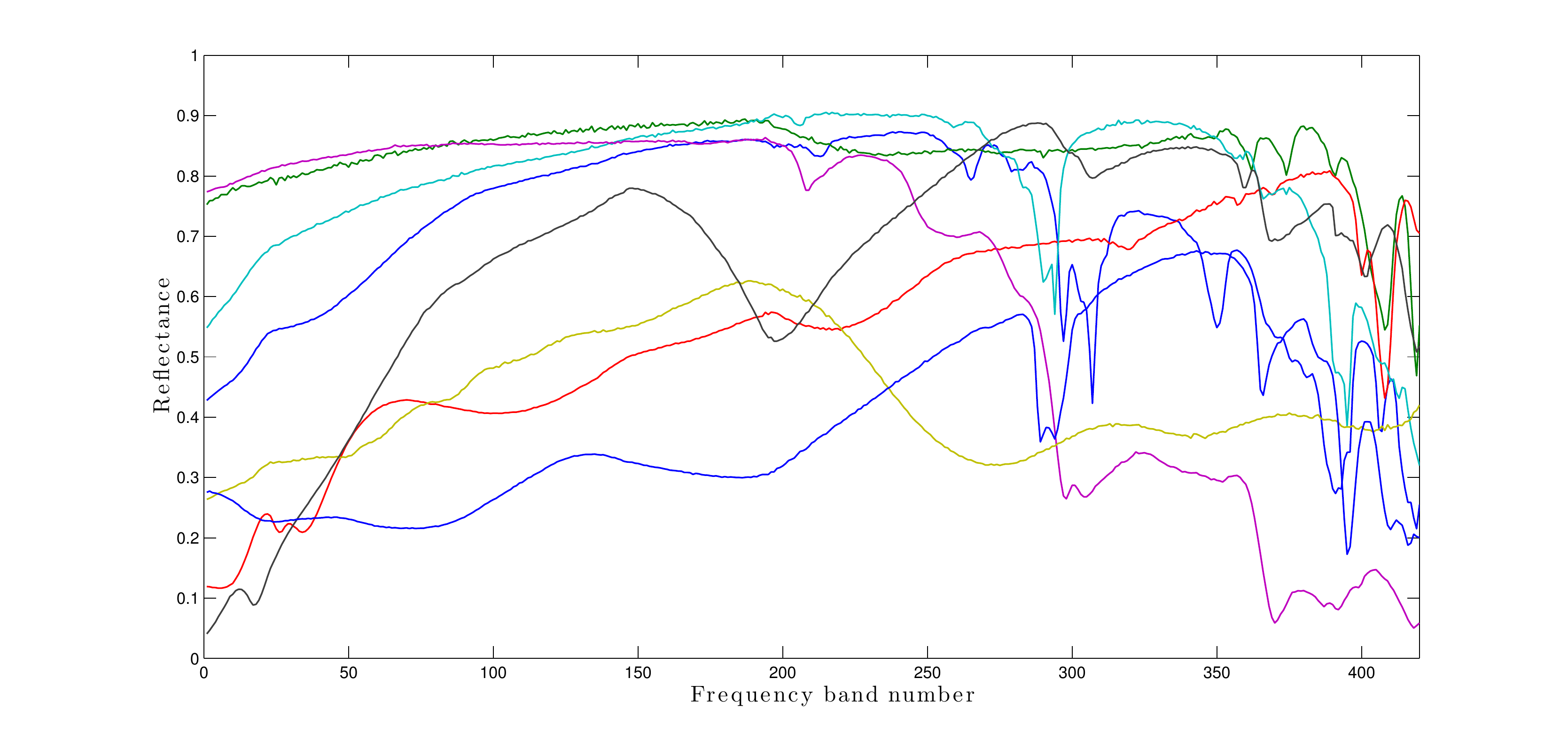}
    \caption{Reflectance of selected endmembers from USGS Library.}
    \label{endmembers}
\end{figure}

The performance of GLUP and NGLUP were evaluated using synthetic data. We used seven endmembers with 420 spectral samples extracted from the USGS library. Figure \ref{endmembers} shows the reflectance of the endmembers. The spectral mutual coherence between two spectra is defined as $\theta_{ij}=\frac{\langle\bs_i,\bs_j\rangle}{\|\bs_i\|\|\bs_j\|}$. The average mutual coherence of the eight endmembers was $\theta_{\text{avg}}=0.9171$. The abundances were generated based on a Dirichlet distribution with unit parameter, as a consequence of which the resulting abundances obeyed the non-negativity and sum-to-one constraint, and were uniformly distributed over this simplex.

\begin{figure}
    \centering
    \includegraphics[scale=0.4]{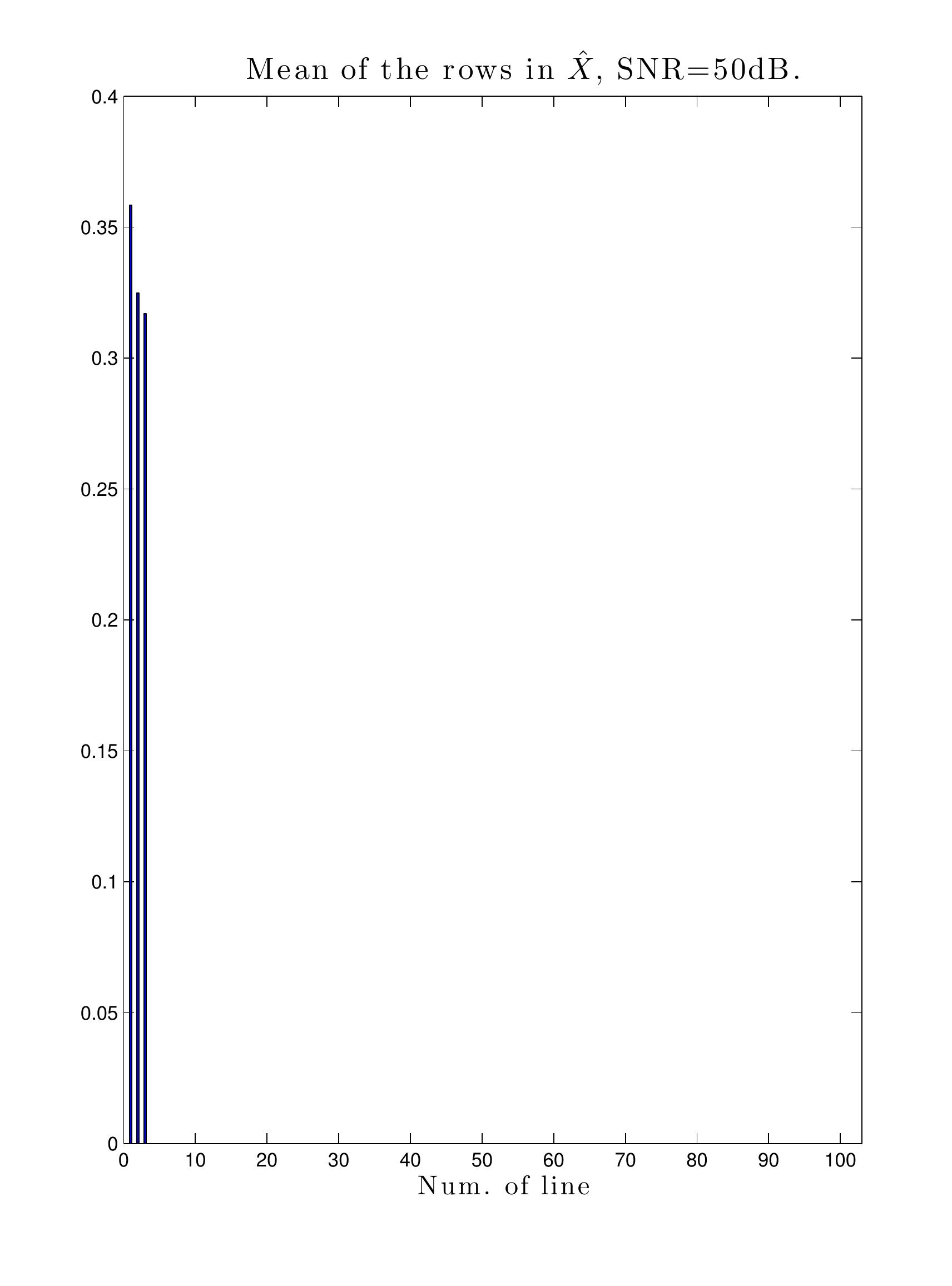}
    \caption{Mean value of each row of $\boldsymbol{\hat{X}}$ estimated with GLUP, obtained with $100$ pixels and $\text{SNR}=50$ dB.}
    \label{GLPCUExample}
\end{figure}

First, we used three endmembers to generate an hyperspectral data set containing $100$ pixels with a SNR of $50$ dB. The pure pixels were indexed by integers $1$--$3$ for simplicity, the mixed pixels being indexed by integers $4$--$100$. We run GLUP algorithm using all the observations ($\bS_{\omega}=\bS$) with $\mu=10$ and $\rho=100$. The primal and dual tolerances were set to $10^{-5}$. Figure \ref{GLPCUExample} shows the mean of each row $\bx_k$ of the estimated abundance matrix $\boldsymbol{\hat{X}}$. We observe that the first three pixels can be identified as the endmembers since the mean values of the first three rows are clearly different from zero. GLUP was able to provide this result in $4.73$ seconds\footnote{Machine specifications: 2.2 GHz Intel Core i7 processor and 8 GB RAM} with a Root Mean Square Error (RMSE), defined as  $\frac{1}{N^2}\|\boldsymbol{\hat{X}}-\bX\|_{\text{F}}^2$, equal to $0.0049$. 

\begin{figure}
    \centering
    \includegraphics[scale=0.4]{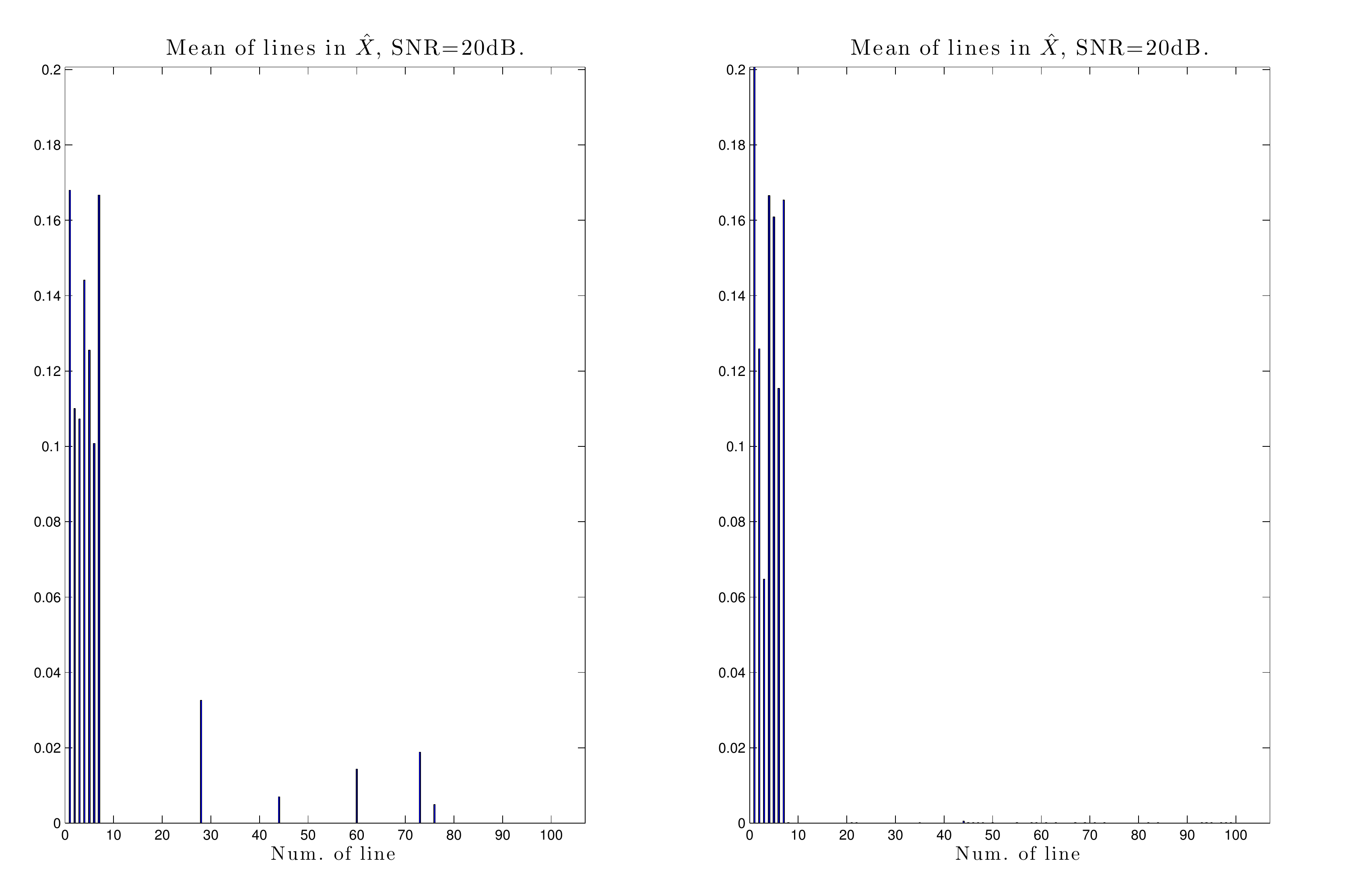}
    \caption{Mean value of each row of $\boldsymbol{\hat{X}}$, obtained with $100$ pixels and $\text{SNR}=20$ dB: GLUP (left), NGLUP (right).}
    \label{MistoNoMisto}
\end{figure}

We tested NGLUP in less favorable conditions by increasing the number of endmembers and decreasing the SNR. To this end, $7$ endmembers were used to generate $93$ mixed pixels. Data were corrupted with an additive Gaussian noise with a SNR of $20$ dB. We tested the algorithm for a maximum number of inner iterations $J_{\text{max}}=1$, $10$ and $100$. We found that NGLUP converged to the same solution even when the number of inner iterations $J$ was equal to $1$. For this reason, only one inner iteration per outer iteration was used for the rest of the experiments. The running time of the algorithm was $45$ seconds. Figure~\ref{MistoNoMisto} shows the mean value of each row $\bx_k$ of the abundance matrix $\hat\bX$ estimated by GLUP and NGLUP algorithms. In both cases, the $7$ largest mean values correspond to the $7$ endmembers. As expected, NGLUP converged to a sparser and more accurate solution than GLUP. 


\begin{table}
\caption{Probability of detecting $\hat{M}$ endmembers, using synthetic data generated with $M=7$ endmembers.}
\begin{center}
\begin{tabular}{  | c | c | c | c | c | c | c | c |   }  
	\hline
	$ M $  & 2 & 3& 6 & 7 & 8  \\  \hline   \hline \nonumber 
	NGLUP &   &   &   &  &     \\ 
	($30$ dB) & 0 & 0 & 0 & $\mathbf{0.98}$ & 0.02  \\ \hline
 	VD  &   &   &   &  &      \\ 
	($30$ dB) &  $\mathbf{0.79}$ & 0.21 & 0 & 0 & 0  \\ \hline  \hline	
	NGLUP &   &   &   &  &    \\ 
	( $20$ dB) & 0 & 0 & 0.01 & $\mathbf{0.96}$ & 0.03 \\ \hline
 	VD  &   &   &   &  &     \\ 
	($20$ dB) &  $\mathbf{1}$ & 0 & 0 & 0 & 0  \\ \hline
\end{tabular}
\label{table:one}
\end{center}
\end{table}

We repeated this simulation $100$ times. For each realization, we examined the number of mean values of the rows of $\boldsymbol{\hat{X}}$ that were larger than a predefined threshold equal to $0.01$. We considered this value $\hat{M}$ as the estimated number of endmembers in the scene. Table \ref{table:one} provides the probability of detecting $\hat{M}$ endmembers with our two approaches, given synthetic data generated with $M=7$ endmembers. The same task was performed using Virtual Dimensionality (VD) \cite{Chang04}. We compared the results of NGLUP with those of VD, the probability of false alarm of VD being set to $10^{-3}$. Table \ref{table:one} shows that NGLUP was able to identify the presence of $7$ endmembers in $98\%$ (resp., $96\%$) of the cases with an SNR of $30$ dB (resp. $20$ dB). VD only identified $2$ endmembers in most cases. Even with higher values of the SNR, VD did not identify the correct number of endmembers. This is due to the fact that VD has asymptotic convergence, and thus requires a very large number of observations in order to converge. This explains the poor performance of VD compared to NGLUP. 

\subsection{Real data}
\label{sec:expReal}

In this section, we shall evaluate the performance of NGLUP using real hyperspectral data. The tests were performed on the so-called images of {Pavia University},\footnote{Available at \text{http://www.ehu.es/ccwintco/index.php/Home}} provided by the ROSIS imaging spectrometer. The scene has a spatial dimension of $610\times 715$, that is, a total of $207,400$ pixels with a spatial resolution of $3.7$ meters per pixel. Each pixel is composed of $102$ spectral samples over the range $430$-$860$ nm.

Given the high dimension of this data set, a subset $\bS$ of $300$ pixels was randomly selected from the available observations. NGLUP was run with $\bS_w=\bS$. Based on this subset of observations, we selected those few that best described the whole scene. The estimated abundance matrix had a few rows different from zero, pointing out the candidate endmembers. Nevertheless, due to the extensive presence of redundant spectra, some rows revealed several occurrences of the same endmember. An additional step was thus required to remove redundant spectra among the endmembers determined by our algorithm. Several procedures have been proposed in the literature to perform this task. For example, in \cite{Esser12}, the authors suggest to use K-means clustering in order to choose a subset of independent observations. In our experiments, we found it sufficient to impose a maximum value of $0.95$ for the mutual coherence among the estimated endmembers. Redundant endmembers according to this criterion were discarded. With Pavia University data, this rule gave us $5$ distinct endmembers.

\begin{figure}

  \centering
  \subfigure[Meadow]{ \includegraphics[scale=0.5]{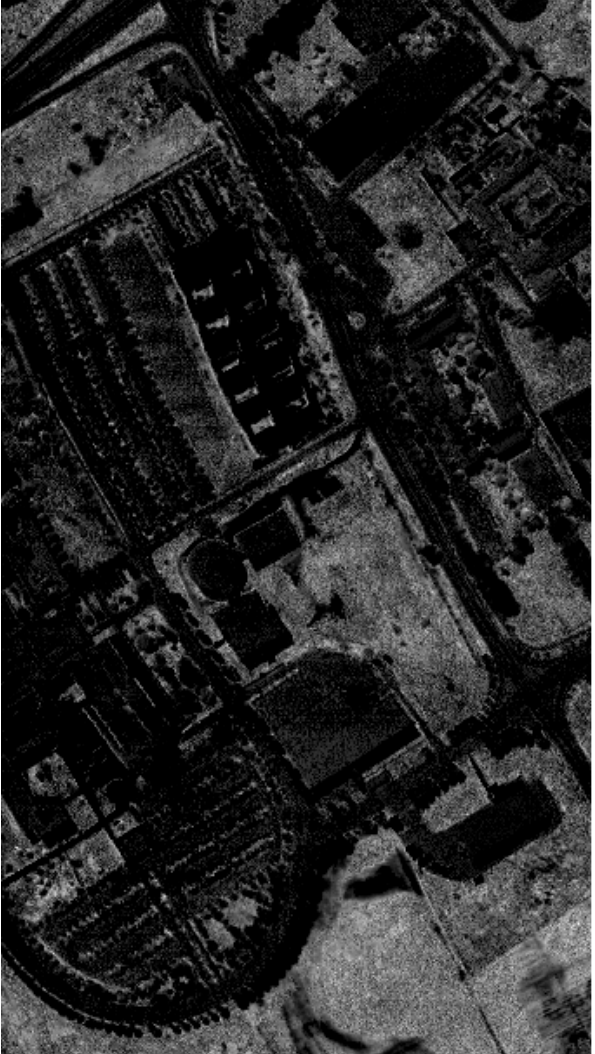}} 
  \subfigure[Tree]{ \includegraphics[scale=0.5]{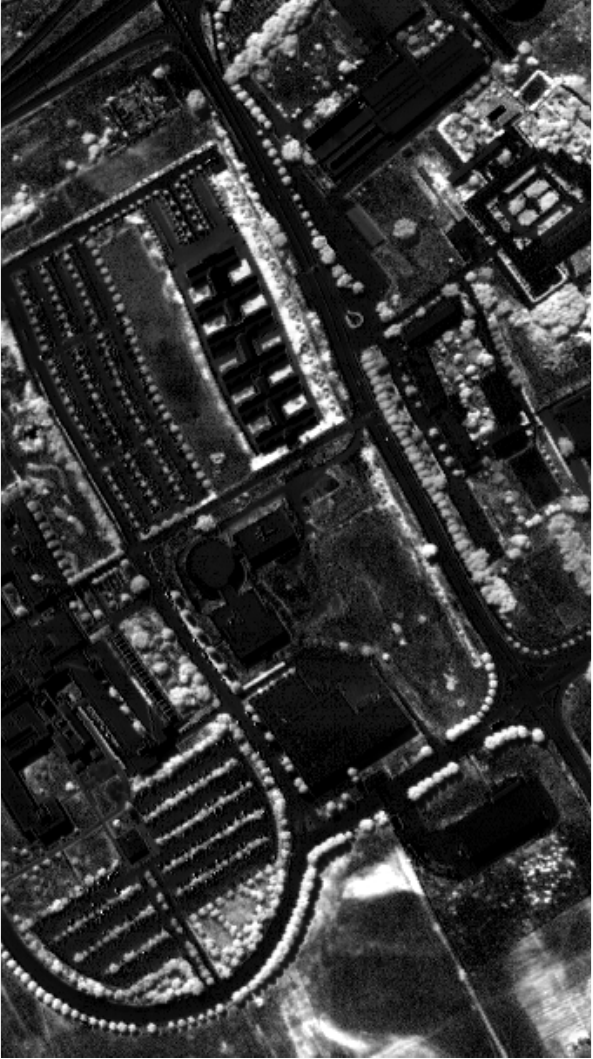}}
  \subfigure[Roof]{ \includegraphics[scale=0.5]{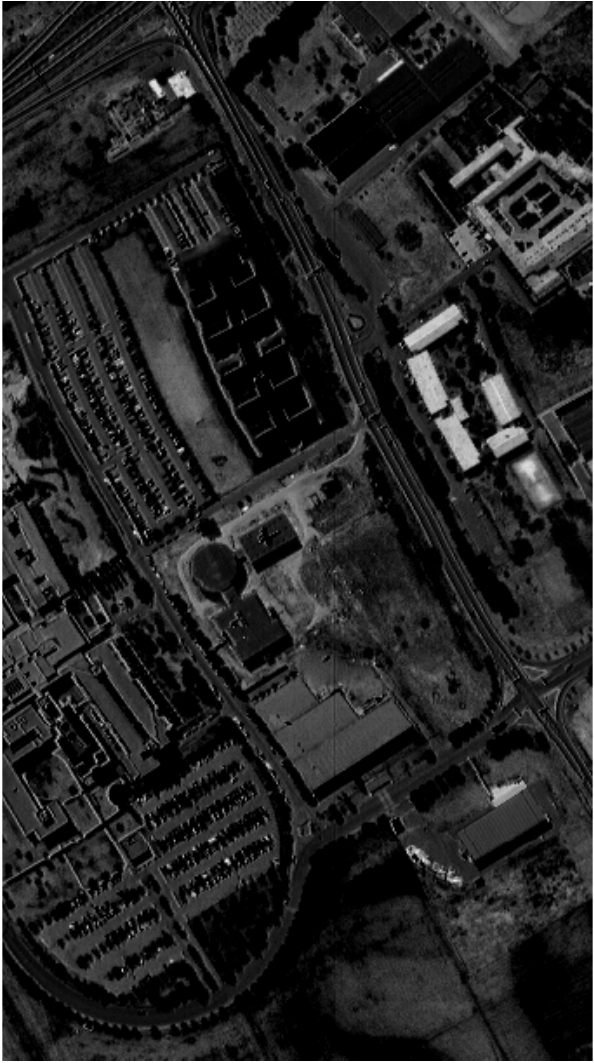}}
  \subfigure[Shadow]{ \includegraphics[scale=0.5]{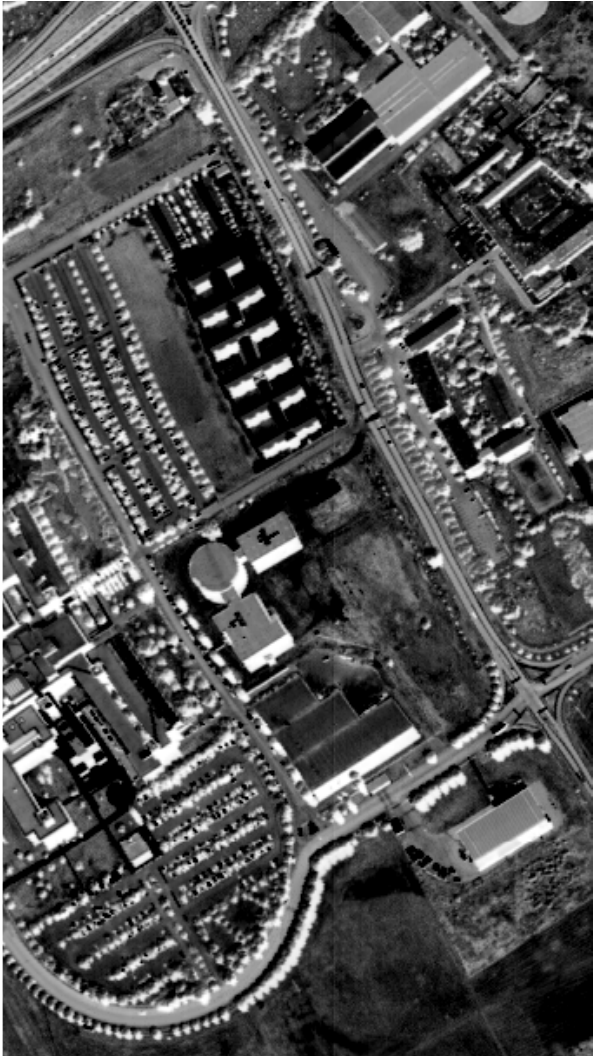}}
  \subfigure[Painted metal]{ \includegraphics[scale=0.5]{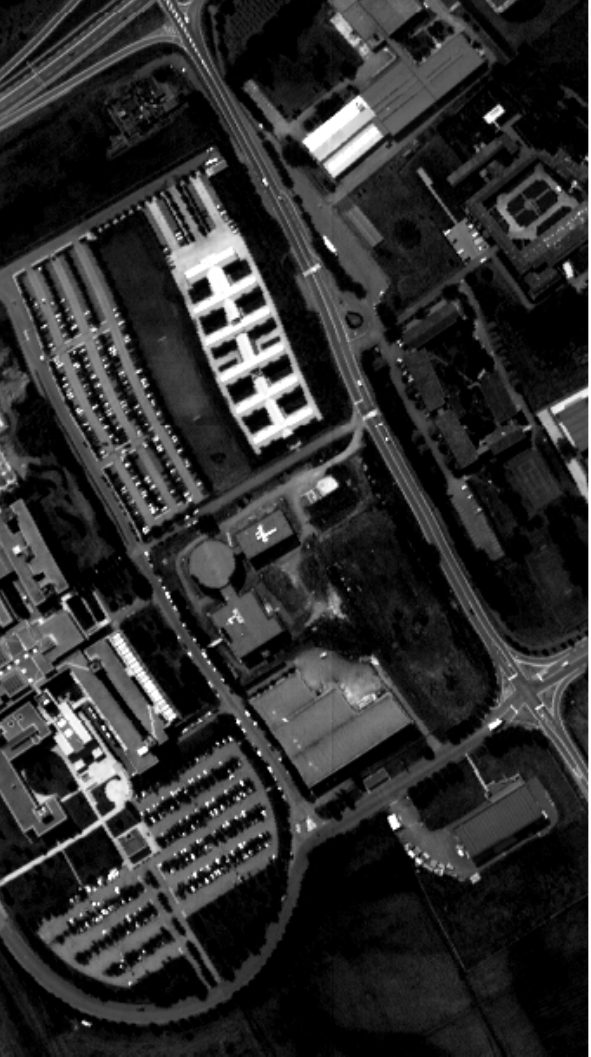}}
	
	\caption{Abundance maps of the five endmembers determined by NGLUP for Pavia University data set.}
	\label{abdmaps}
\end{figure}

We assumed that these endmembers, obtained from a small subset of the observations, were valid for the whole scene. This assumption can be justified by the fact that the image has lots of homogeneous surfaces where the spectral variability is negligible. We then used these endmembers and applied the Fully Constrained Least Squares (FCLS) algorithm on the whole data set. Figure \ref{abdmaps} shows the abundance map for every endmember determined by NGLUP. The maps successfully describe the urban features of the scene and highlight its topography. They cast the pixels as combinations of meadow, tree, shadow, roof, painted metal sheet (with asphalt). Finally, we compared the performance of NGLUP with N-FINDR. Using the same subset of observations as NGLUP, we determined $5$ endmembers with N-FINDR. Then, we applied FCLS on the whole image with the endmembers determined by N-FINDR. Table \ref{table:two} shows the RMSE, the maximum and average spectral angles obtained for both methods. This comparison shows that NGLUP outperformed N-FINDR.


\begin{table}
\caption{Performance of NGLUP and N-FINDR for Pavia University data set}
\begin{center}
\begin{tabular}{  | c | c  | c | c |  }  
	\hline
	 Algorithm & RMSE     &  max angle (rad) & avg angle (rad) \\  \hline   \hline \nonumber 
	 NGLUP     & $0.0287$ &    $0.7468$ & $0.074$      \\ \hline
	 N-FINDR    & $0.0641$ &     $1.0592$ & $0.1549$     \\ \hline
\end{tabular}
\label{table:two}
\end{center}
\end{table}

\section{Conclusion and perspectives}
\label{sec:conclusion}
In this work, we presented two approaches for blind and fully constrained unmixing. Both methods are based on mixing models with increasing complexity, and allow to simultaneously determine the endmembers and estimate their local abundance in the scene. Compared to the first model called GLUP, the second model NGLUP explicitly considers that endmembers present in the scene are corrupted by noise. Experiments on synthetic and real data demonstrated the excellent performance of both approaches. Future work includes their extension to an online framework, which would allow to reduce their complexity and to make them adaptive to changing environmental conditions.

\appendix
\begin{IEEEproof}
Since problem \eqref{Pmisto} is convex, we simply have to check the validity of the solution in the two cases $\|(\bv )_+\|_2 > \alpha$ and $\|(\bv )_+\|_2 < \alpha$.
Let $f_0(\bz )=\frac{1}{2} \| \bz -\bv \|_2^2 + \alpha  \| \bz \|_2$. For $\|(\bv )_+\|_2 > \alpha$, the gradient of $f_0$ is given by
\begin{equation}
\nabla f_0(\bz ^\ast) = \left( 1 + \frac{\alpha}{\|\bz ^\ast\|_2}\right)\bz ^\ast - \bv 
\end{equation}
Replacing by the appropriate expression from \eqref{Pmisto} yields
\begin{align}
&\nabla f_0(\bz ^\ast) = (\bv )_+ - \bv  \geq 0 \\
&\bz ^\ast_i \cdot \nabla f_0(\bz ^\ast)_i  \propto  ((\bv )_+)_i \cdot ((\bv )_+ - \bv )_i = 0
\end{align}
These two conditions correspond the optimality conditions, which means that $\bz \succeq \bzero$ is a solution for the constrained problem. For more details, refer to section 4.2.3 in \cite{Boyd08}.

For the second case, note that for every $\bz \succeq \bzero$, we have
\begin{equation}
	\sum_i \bz _i\bv _i \leq \sum_i \bz _i(\bv _i)_+ \leq \|\bz \|_2 \cdot \|(\bv )_+\|_2
\end{equation}
It follows that
\begin{equation}
\begin{split}
f_0(\bz )-f_0(\bzero) & = \frac{1}{2}\sum_i \bz _i^2 - \sum_i \bz _i\bv _i +\alpha \|\bz \|_2   \\
& \geq  \frac{1}{2}\|\bz \|_2^2 -\|\bz \|_2 \cdot \|(\bv )_+\|_2  +\alpha \|\bz \|_2   \\
& \geq \frac{1}{2}\|\bz \|_2^2 +\|\bz \|_2 ( \alpha - \|(\bv )_+\|_2 ) \label{major1}
\end{split}
\end{equation}

This proves that for $\|(\bv )_+\|_2 \leq   \alpha $, the minimum is reached for $\bz ^\ast = \bzero$.
\end{IEEEproof}

\bibliographystyle{IEEEbib}
\bibliography{ref}

\end{document}